\documentclass[twocolumn,showpacs]{revtex4}%
\UseRawInputEncoding
\usepackage{amssymb}
\usepackage{units}
\usepackage{amsmath}
\usepackage{graphicx}
\usepackage{epstopdf}
\usepackage{dcolumn}
\usepackage{bm}
\usepackage{xcolor}
\usepackage{xfrac}
\usepackage{amsfonts}%
\usepackage{tikz}
\usepackage{ulem}
\begin{document}
\title{Optically detected flip-flops between different spin ensembles in diamond}
\author{Sergei Masis}
\affiliation{Andrew and Erna Viterbi Department of Electrical Engineering, Technion, Haifa
32000 Israel}
\author{Sergey Hazanov}
\affiliation{Andrew and Erna Viterbi Department of Electrical Engineering, Technion, Haifa
32000 Israel}
\author{Nir Alfasi}
\affiliation{Andrew and Erna Viterbi Department of Electrical Engineering, Technion, Haifa
32000 Israel}
\author{Oleg Shtempluck}
\affiliation{Andrew and Erna Viterbi Department of Electrical Engineering, Technion, Haifa
32000 Israel}
\author{Eyal Buks}
\affiliation{Andrew and Erna Viterbi Department of Electrical Engineering, Technion, Haifa
32000 Israel}
\date{\today }

\begin{abstract}
We employ the technique of optical detection of magnetic resonance to study dipolar interaction in diamond between nitrogen-vacancy color centers of different crystallographic orientations and substitutional nitrogen defects. We demonstrate optical measurements of resonant spin flips-flips (second Larmor line), and flip-flops between different spin ensembles in diamond. In addition, the strain coupling between the nitrogen-vacancy color centers and bulk acoustic modes is studied using optical detection. Our findings may help optimizing cross polarization protocols, which, in turn, may allow improving the sensitivity of diamond-based detectors.
\end{abstract}
\maketitle

\section{Introduction}
\label{sec:Int}
\newcommand{\NVm}{NV$^-$}
\newcommand{\NVo}{NV$^0$}
\newcommand{\mS}{m_\mathrm{S}}
A color center in diamond composed of a substitutional nitrogen and a vacancy in the crystal lattice (NV)~\cite{Doherty_1} draws lately a considerable attention, particularly in the negatively charged state (\NVm). The \NVm electronic spin state can be polarized and read out with light.
Dense ensembles of \NVm centers were demonstrated to be applicable for magnetometry~\cite{Rondin_056503}, classical~\cite{Dhomkar_e1600911} and quantum \cite{Zhu2011,Kubo2011,Amsuss2011,Grezes_021049} information storage, and recently, a maser implementation \cite{Breeze_493}.

Dipolar coupling affecting a given \NVm in a diamond crystal is commonly dominated by other \NVm s, having either parallel or non-parallel lattice orientation, and by substitutional nitrogen (P1) defects, having a density typically higher by at least an order of magnitude. The interaction between the ensembles might be a quantum resource~\cite{Bermudez_150503} or a source of decoherence~\cite{BarGill_1}. 

Spontaneous, i.e. phonon assisted, spin flip-flops between ensembles (when one of the spins changes from high to low energy states, and the other changes in the opposite direction), were demonstrated before~\cite{Bloembergen_445,Alfasi_214111,Armstrong_1569,Wang_4135}. In this work we demonstrate direct and unambiguous stimulated flip-flop interaction between the parallel and the non-parallel \NVm ensembles or the P1 ensemble. A higher order process involving five spins interaction is reported elsewhere~\cite{Masis_013852}.
In addition, we observe dipolar spin-spin interaction~\cite{Waller_370} at double the Larmor frequency inside a single \NVm ensemble parallel to the external magnetic field. The increased sensitivity of our setup is attributed to a combination of high \NVm density, low temperature and lock-in amplification.

In the context of sensing, our results may be described as optically detected electron-electron double resonance~\cite{dorio2012multiple} (ELDOR, also known as DEER). Unlike the traditional, microwave (MW) cavity based ELDOR, optical detection allows increased sensitivity, wide range of MW frequencies and higher spacial resolution. In the context of hyperpolarization~\cite{Belthangady_157601}, our method is closely related to dynamic nuclear polarization~\cite{Kamp2018} (DNP), where in the role of slow nuclear spins are the electronic spins of the P1, and in the role of fast electron spins are the optically polarized \NVm.
Hyperpolarized P1 then might serve as a low noise spin bath for the \NVm, or as a target spin ensemble on its own right, e.g. for applications of P1 maser.
\section{ODMR measurements}
\label{sec:ODMR}
\begin{figure}
\includegraphics[width=\columnwidth]{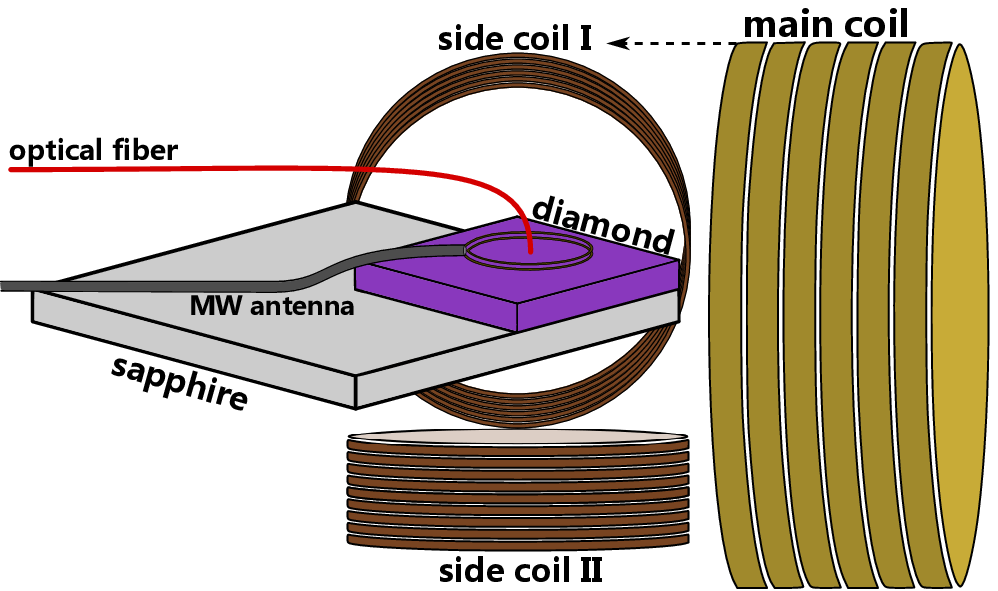}
\begin{picture}(0,0){
    \setlength{\unitlength}{1cm}
    \put(-4,0){
        \frame{
            \includegraphics
                 {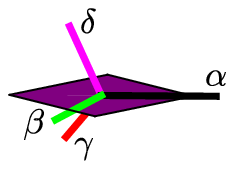}
            }
        }
}
\end{picture}
\caption{
Experimental setup. Radio frequency radiation is introduced by a flexible coaxial cable with a MW loop antenna termination pressed against the $(110)$ diamond face. A multi-mode optical fiber normal to the diamond wafer, and positioned in the center of the loop antenna, is used both to illuminate and to collect PL. A superconducting solenoid (main coil) produces strong longitudinal magnetic field $B_\parallel$, aligned in the direction of diamond $\alpha$ vector by two smaller superconducting coils. Inset: diamond crystallographic directions. Vectors $\alpha(\left[1\bar1\bar1\right])$ and $\beta(\left[\bar11\bar1\right])$ are in the $(110)$ plane, nominally parallel to the substrate surface depicted by the purple rectangular patch. Vectors $\gamma(\left[\bar1\bar11\right])$ and $\delta(\left[111\right])$ are not in the $(110)$ surface and are mirrored by it.
}
\label{FigSetup}
\end{figure}
\begin{figure*}
\begin{center}
\includegraphics {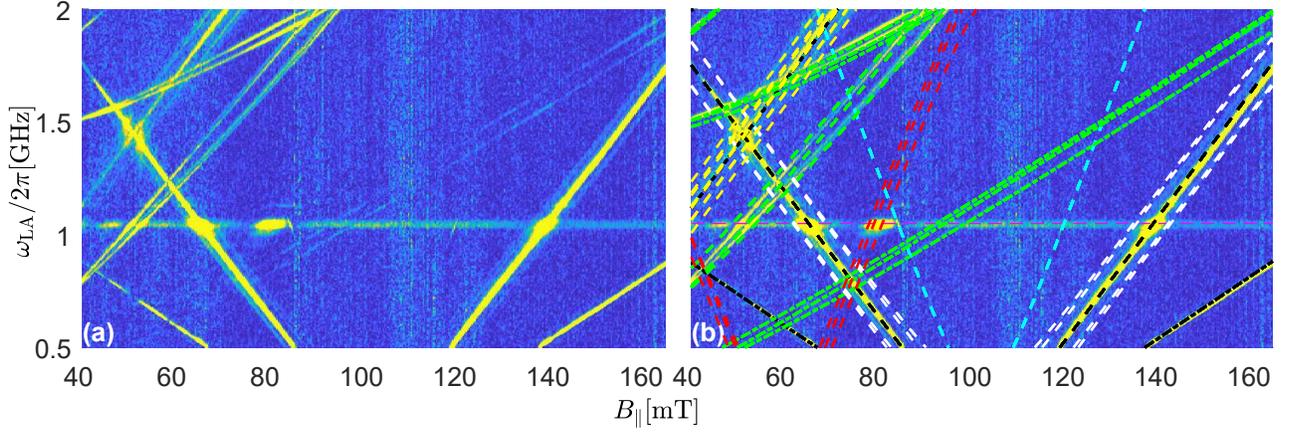}
\caption{Big picture ODMR plot. Side coils were tuned to compensate the misalignment at $B_\parallel=38~\unit{mT}$. Dashed black lines (color online) correspond to $T^{-1,0}_\alpha$ transition, symmetrical about the $\mS=-1$ and $\mS=0$ levels anticrossing at $102~\unit{mT}$. White lines are due to the hyperfine interaction of \NVm with first shell 13C isotope~\cite{Alfasi_214111,Felton_075203,Mizuochi_041201,Simanovskaia_224106,Kamp_045204,Loubser_1201}. Black dash-dotted lines represent multiphoton processes at half the frequency of $T^{-1,0}_\alpha$. Cyan line, only visible in the data at the vicinity of the MW resonator frequency, represents the double Larmor line of $T^{-1,0}_\alpha$, interpreted as flip-flip (see Fig.~\ref{Fig:LAC}). Green dashed lines represent the $T^{-1,+1}_{\beta,\gamma,\delta}$ transitions of the three \NVm vectors not parallel to the bias magnetic field. The top (bottom) dash-dotted green lines represent the double-photon transitions $T^{-1,0}_{\beta,\gamma,\delta}$($T^{-1,+1}_{\beta,\gamma,\delta}$). The red lines (visible only near the resonator resonance) correspond to the difference between the transitions $T^{-1,+1}_{\beta,\gamma,\delta}$ and $T^{-1,0}_\alpha$, interpreted as spin flip-flop (see Fig.~\ref{Fig:NVNV}). 
The yellow lines correspond to the P1 hyperfine-split $\mS=-1/2$ to $\mS=+1/2$ transition (see Fig.~\ref{Fig:P1}), while the parallel black dotted line in their center represents the double-photon $T^{-1,+1}_\alpha$ transition.
The magenta line of the MW resonator is of unclear origin~\cite{Alfasi_063808}, however the signal enhancement of other resonances in the vicinity of the resonator frequency might be attributed to a higher effective MW power.
} 
\label{Fig:resODMR}
\end{center}
\end{figure*}

\newcommand{\w}[1][LA]{\omega_\mathrm{{#1}}}
\newcommand{\HNV}{\mathcal{H}_{\mathrm{NV}}}
\newcommand{\DNV}{D_\mathrm{NV}}
\newcommand{\ENV}{E_\mathrm{NV}}
\newcommand{\gel}{\gamma_\mathrm{e}}

When hyperfine interaction is disregarded, the \NVm ground state spin triplet Hamiltonian becomes  \cite{Ovartchaiyapong_1403_4173,MacQuarrie_227602}
\begin{equation}
\frac{\HNV}{\hbar}=
\frac{\DNV S_z^2}{\hbar^2}
+\frac{\ENV\left( S_+^2+S_-^2\right)}
    {2\hbar^{2}}
-\gel\frac{\mathbf{B}\cdot\mathbf{S}}{\hbar}\;,
\label{H_NV TGS}
\end{equation}
where $\DNV=2\pi\times2.87~\unit{GHz}$ is the zero field splitting, $\ENV\ll\DNV$ is a strain-induced splitting, $\gel=2\pi\times28.03~\unit{GHz}\unit{T}^{-1}$ is the electron spin gyromagnetic ratio, $\mathbf{B}$ is the applied magnetic field,
$\mathbf{S}=S_{x}\mathbf{\hat{x}}+S_{y}\mathbf{\hat{y}}+S_{z}\mathbf{\hat{z}}$
is the spin $S=1$ angular momentum vector operator and $S_{\pm}=S_{x}\pm iS_{y}$. Under continuous laser excitation, \NVm is polarized to the spin state having magnetic quantum number $\mS=0$, which has a slightly brighter photoluminescence (PL) \cite{Doherty_1}. Introducing MW irradiation at frequency $\w/2\pi$ resonant with $\mS=0$ to $\mS=\pm1$ transition reduces the spin polarization and PL, allowing optical detection of the magnetic resonance (ODMR). We will use a nomenclature of $T_v^{i,j}$ to indicate a resonant transition in \NVm center parallel to vector $v\in\{\alpha,\beta,\gamma,\delta\}$ (see inset in Fig.~\ref{FigSetup}) from state $i$ to state $j$, where $i,j\in\{-1,0,+1\}$. Both $i$ and $j$ states are eigenstates of the Hamiltonian $\HNV$ (\ref{H_NV TGS}), where $i$ and $j$ indicate the magnetic quantum number $\mS$ when $[\HNV,S_{z}]=0$ (i.e. when the externally applied magnetic field $\mathbf{B}$ is parallel to the symmetry axis of the \NVm center, and strain-induced splitting is disregarded). A typical ODMR plot is shown in Fig.~\ref{Fig:resODMR}.

Our ODMR plots show not only the \NVm resonances, but also other defects. Among them - nitrogen 14 (nuclear spin 1) substitution defect (P1) in diamond \cite{Cook_99,Loubser_1201}, which has four locally stable configurations. In each configuration a static Jahn-Teller distortion \cite{Smith_1546} occurs, and an unpaired electron is shared by the nitrogen atom and by one of the four neighboring carbon atoms, which are positioned along one of the
lattice directions $\left\langle 111\right\rangle$ \cite{Takahashi_047601,Hanson_087601,Hanson_352,Wang_4135,Broadway_1607_04006,Shim_1307_0257,Smeltzer_025021,Shin_205202,Clevenson_021401,Schuster_140501}. The transition frequencies are calculated (see Figs.~\ref{Fig:resODMR}, \ref{Fig:NVNV} and \ref{Fig:P1}) by numerically diagonalizing the P1 spin Hamiltonian $\mathcal{H}_{\mathrm{P}1}$ \cite{Cox_551,Alfasi_214111} 
using the following parameters:
nitrogen 14 nuclear gyromagnetic ratio $\gamma_{\mathrm{n}}=2\pi\times3.0766
\operatorname{MHz}$, nitrogen 14 quadrupole coupling $Q_{\mathrm{P}1}=-2\pi\times3.97\operatorname{MHz}$, longitudinal hyperfine coupling $A_{\mathrm{P}1,\parallel}=2\pi\times114\operatorname{MHz}$ and transverse hyperfine coupling $A_{\mathrm{P}1,\parallel}=2\pi\times81.3\operatorname{MHz}$.

\begin{figure}
\begin{center}
\includegraphics[width=\columnwidth] {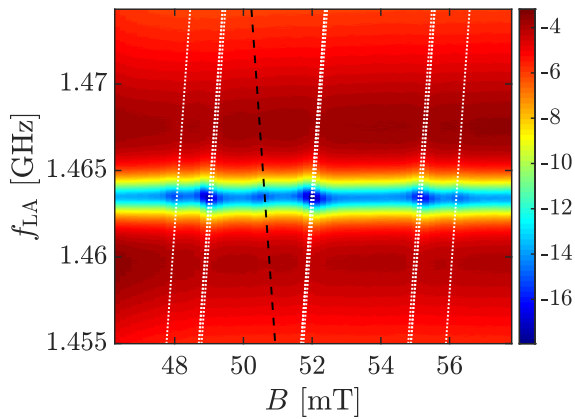}
\caption{
MW cavity ESR measurement without laser polarisation, as reported in reference~\cite{Alfasi_214111}, with 5 resolved P1 resonances fit by white dotted lines [see Eq.~(3) in~\cite{Alfasi_214111}] and the \NVm parallel to magnetic field $\mS=0$ to $\mS=-1$ resonance fit by a black dashed line [see Eq.~(1) in~\cite{Alfasi_214111}]. The color represents reflection coefficient $S_{11}$ in dB units, and the superconducting resonator is different, having a frequency of $1.463~\unit{GHz}$. The diamond orientation is similar to the one in the current work, and the  fit parameters are as in caption of Fig.2. of~\cite{Alfasi_214111}. The comparable cavity frequency shift as result of \NVm resonance ($1/4$ of total \NVm ensemble) and the weakest of P1 resonances ($1/12$ of the P1 ensemble) results with an estimate of \NVm to P1 concentration ratio of 1:3. 
} 
\label{Fig:ESR}
\end{center}
\end{figure}

Type Ib HPHT single crystal $[110]$ grown diamond with $<200~\unit{ppm}$ nitrogen  concentration (same as in~\cite{Alfasi_214111}) was laser-cut, polished, irradiated with $2.8~\unit{MeV}$ electrons at a doze of $8\times10^{18}~\unit{e/cm^2}$, annealed for $2~\unit{h}$ at $900~\unit{C}$ and boiled for $1~\unit{h}$ in equal mixture of Perchloric, Sulfuric and Fuming Nitric acids, resulting with a fluorescent-count measured \NVm concentration of $n_\mathrm{S}=3.25\times10^{17}~\unit{cm^{-3}}$. The \NVm:P1 ratio is estimated from electron spin resonance (ESR) data to be $1:3$ (see Fig.~\ref{Fig:ESR})~\cite{Alfasi_214111}. The diamond is oriented at room temperature to have the $\alpha([1\bar1\bar1])$ axis coinciding with the main coil axis and the (110) surface orthogonal to the 2nd coil.  The diamond wafer is glued to a sapphire substrate carrying a $1.05~\unit{GHz}$ superconducting spiral resonator (see Fig.~\ref{FigSetup}). The resonator is not used in this experiment. All measurements are performed at $3.5~\unit{K}$. Throughout the paper, the diamond misalignment is fit by azimuthal  $\theta=2.86^\circ$ and polar $\phi=1.71^\circ$ angles with respect to the coordinate system defined by the axial directions of the 1st, 2nd and the main superconducting coils (see Fig.~\ref{FigSetup}). A $637~\unit{nm}$ red laser is used to excite the NV centers at the \NVm zero-phonon line (ZPL), rather than the more commonly used $532~\unit{nm}$, in order to reduce sample heating. Despite the expected photoionization to the neutrally charged NV state, no trace of these defects was recorded in the ODMR scans. The returning PL light passes through a beam splitter and a $700~\unit{nm}$ long pass filter before measurement with a reverse-biased photo-diode (PD). MW loop antenna is connected directly to a synthesizer, $100\%$ amplitude modulated by a $151~\unit{Hz}$ sine wave. The PD signal is demodulated by a lock-in amplifier~\cite{Anishchik_115142}, and the resulting amplitude is recorded. In all the two dimensional scans the magnetic field is scanned first, slowly enough to avoid hysteresis.

\section{Same Spin-Spin Interaction}
\label{sec:H2}
\begin{figure}
\begin{center}
\includegraphics {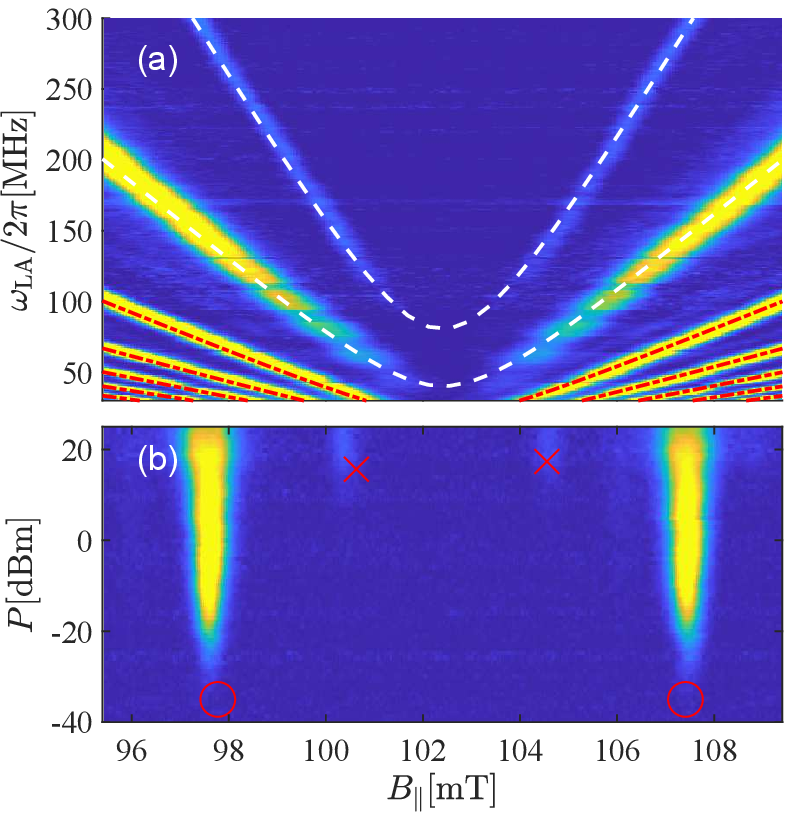}
\begin{picture}(0,0){
    \setlength{\unitlength}{1cm}
    \put(-4.5,6.5){
        \frame{
            \includegraphics[width=1.5cm]             {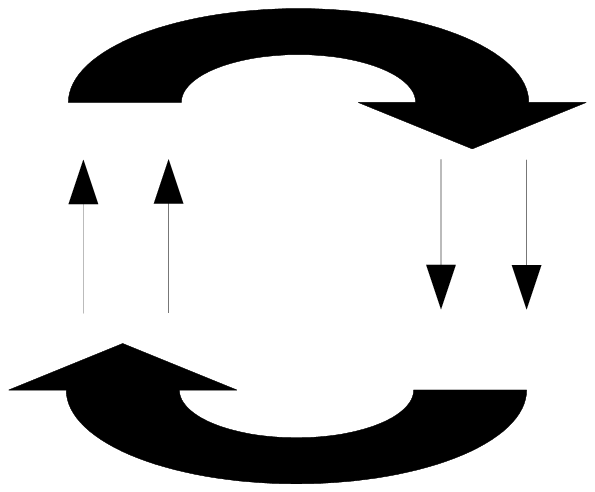}
            }
        }
}
\end{picture}
\caption{Double Larmor frequency ODMR. \textbf{(a)} White dashed lines represent fit of first and second Larmor frequencies calculated according to Eq.~(\ref{eq:wa}) with $\psi=0.8^\circ$, incorporating both the diamond misalignment and side coils correction. The red dash-dotted lines are calculated as multi-photon resonances by Eq.~(\ref{omega_BS,l}). Alternatively, the $1/m$ hyperbolas might be attributed to higher harmonics of the synthesizer, enhanced by amplitude modulation. Inset: schematic representation of spin flip-flip process. \textbf{(b)} MW power dependency of the ODMR signal at $\w=2\pi\times145~\unit{MHz}$. Circles mark the beginning of first hyperbola visibility ($P_1$), indicating $2\%$ of the maximal recorded signal. Crosses mark the expected beginning of the second hyperbola visibility, given by $R_2P_1$ [see Eq.~(\ref{eq:Anderson})].
 The signal reduction at high MW power is due to lock-in saturation artifact; throughout our measurements higher laser and MW powers lead to higher ODMR contrast.
} 
\label{Fig:LAC}
\end{center}
\end{figure}
For \NVm aligned closely along the magnetic field direction (having a misalignment angle $|\psi|\ll1~\unit{rad}$), a ground state levels anticrossing (GSLAC) between the states $\mS=0$ and $\mS=-1$ occurs at $\DNV/\gel=102.4~\unit{mT}$. The resonance angular frequency $\w[a]$ and the effective transverse drive amplitude $\w[t]$ are given~\cite{Masis_013852,Masis_PhDThesis} by 
\begin{eqnarray}
    \w[a]=\w[a0]\sqrt{1+\eta^2}~,
    \label{eq:wa}
\end{eqnarray}
\begin{eqnarray}
    \w[t]=\w[LA1]\eta/\sqrt{1+\eta^2}~,
    \label{eq:wt}
\end{eqnarray} where $\w[a0]=\sqrt{2}\DNV|\psi|$, $\w[LA1]$ is the MW loop antenna drive amplitude, and the dimensionless detuning parameter $\eta$ is given by $\eta = ( \DNV-\gel B) /\w[a0]$. The signal is recognizable in the ODMR scan as a hyperbola disappearing near the GSLAC, due to spin states mixing and subsequent inefficient MW depolarization required for the ODMR contrast. 

Dipolar coupling gives rise to an absorption resonance at an angular frequency $2\w[a]$ [second hyperbola, see Fig.~\ref{Fig:LAC}(a)], due to spins flip-flip. Fractional hyperbolas at angular frequencies $\w[a]/l$ for integer $l$ are attributed to multi-photon processes or higher harmonics in the excitation signal. The Bloch-Siegert shift corresponding to the $l$'th multi-photon processes is given by [see Eq.(6.378) of ~\cite{Buks_QMLN}]

\begin{equation}
\omega_{\mathrm{BS},l}=\sum_{l^{\prime}\neq l}\frac{\w[t]^{2}\left(
J_{l^{\prime}}\left(  \frac{\omega_{\mathrm{S}1}}{\w[LA]}\right)
+J_{l^{\prime}+2}\left(  \frac{\omega_{\mathrm{S}1}}{\w[LA]}\right)  \right)
^{2}}{2\left(  l-l^{\prime}\right)  \w[LA]}\;, \label{omega_BS,l}%
\end{equation}
where $\omega_{\mathrm{S}1}$ is longitudinal excitation amplitudes. The shift is too small to be resolved in our measurements, hence we conclude that $\w[LA1]<1~\unit{MHz}$, and the ODMR linewidth is caused not by power broadening, but by increased \NVm transverse relaxation or by inhomogeneous broadening.

The ratio between the second and first hyperbolas strength at the effective magnetic field $B_\mathrm{a}=\w[a]/\gamma_\mathrm{e}$ is given by~\cite{Anderson_1517,Cheng1961,Broer_801,Daycock1969} 
\begin{equation}
    R_2=\frac{2}{3}\langle(\Delta B)^2\rangle_{av}/B_\mathrm{a}^2\,,
\label{eq:Anderson}
\end{equation}
where, for disordered defects location, the azimuthally averaged local field is  $\langle(\Delta B)^2\rangle_{av}=\pi\mu_0^2\hbar^2\gamma^2_\mathrm{e}S(S+1)\Sigma_j r_{ij}^{-6}$ and $r_{ij}$ is the distance between defects $i$ and $j$. 
For diluted diamond lattice $\Sigma_j r_{ij}^{-6}=r_\mathrm{eff}^{-6}$, where we calculate $r_\mathrm{eff}=0.907/\sqrt[3]{\rho_\mathrm{NV}}$ and $\rho_\mathrm{NV}=3.25/4\times10^{17}~\unit{cm}^{-3}$ is the density of \NVm parallel to one of the crystallographic vectors.
A quantitative comparison of first and second hyperbolas visibility is given in Fig.~\ref{Fig:LAC}(b), and a good agreement to optically determined NV concentration is found.

\section{Different Spin-Spin Interaction}
\label{sec:flipflop}
\begin{figure*}
\begin{center}
\includegraphics {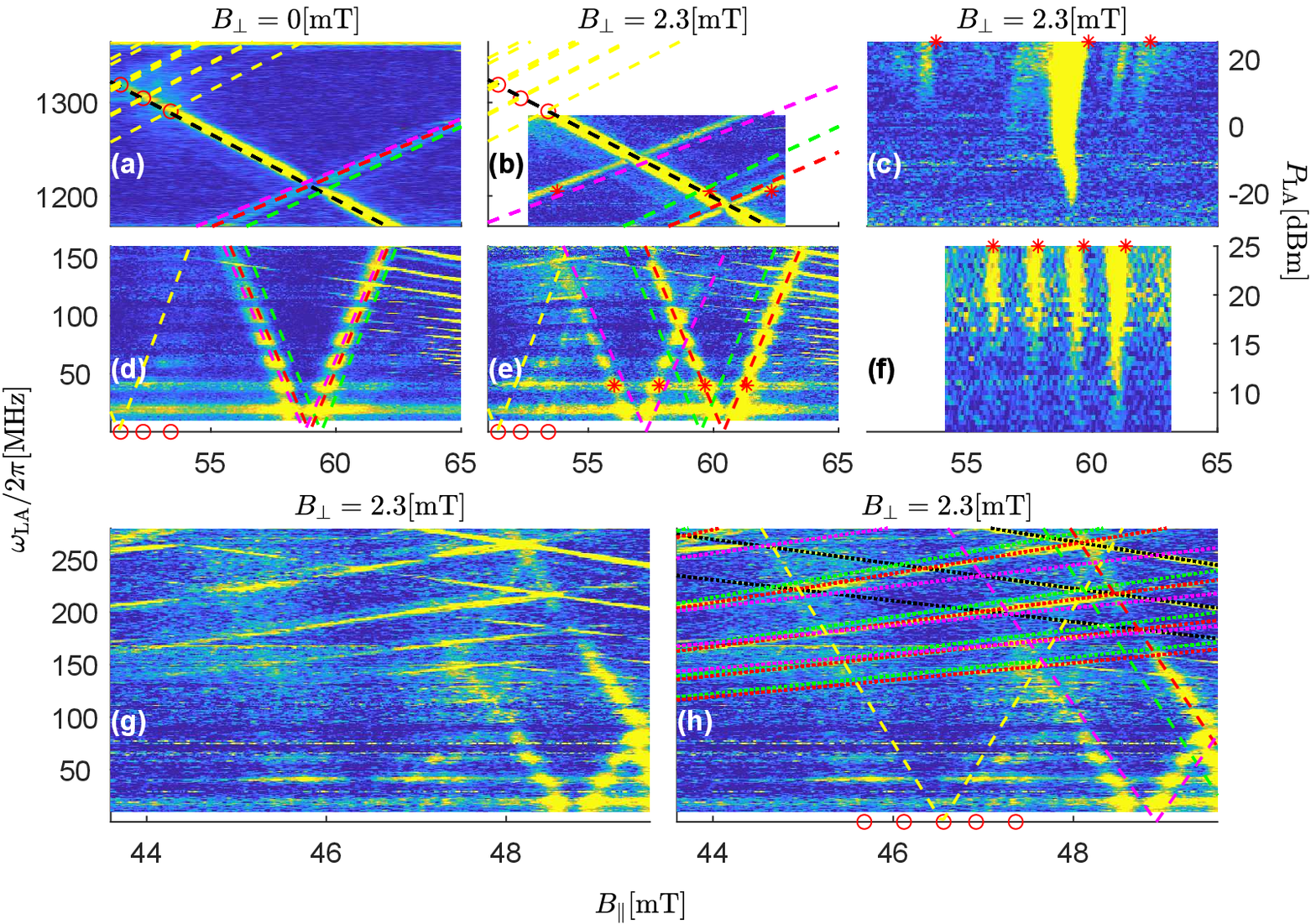}
\begin{picture}(0,0){
    \setlength{\unitlength}{1cm}
    \put(-12.8,3.5){
        \frame{
            \includegraphics[width=1.5cm]             {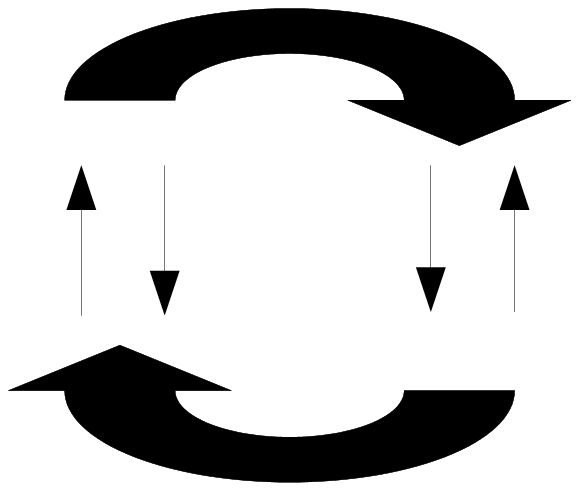}
            }
        }
}
\end{picture}
\caption{
Flip-flop ODMR.
\textbf{(a)} Black/green/red/magenta (color online) are the 
$T^{-1,0}_{\alpha}$/
$T^{-1,+1}_{\beta}$/
$T^{-1,+1}_{\gamma}$/
$T^{-1,+1}_{\delta}$ 
transitions calculated according to Eq.~(\ref{H_NV TGS}). Yellow lines are P1 hyperfine transitions calculated as in Fig.~\ref{Fig:P1}. The side coils magnetic field is set to compensate the sample misalignment at the main coil field of $B_\parallel=59~\unit{mT}$, as validated by measuring the degeneracy of $T^{-1,0}_{\beta,\gamma,\delta}$ transitions (not shown).
\textbf{(b)} Same as (a), but the 2nd coil magnetic field is increased by $B_\perp=2.3~\unit{mT}$, lifting the degeneracy of the \NVm parallel to $\beta,\gamma,\delta$ vectors.  
\textbf{(c)} MW power scan along the $1210~\unit{MHz}$ line in the otherwise similar conditions to (b). Red asterisks in (b) and (c) mark similar points in the parameter space. 
\textbf{(d)} Taken in the same magnetic conditions as (a). Green/red/magenta lines are the difference between the corresponding color lines and the black line in (a).  Yellow line is the difference between the P1 $\mS=+1/2$ to $\mS=-1/2$ transition with nuclear spin number $m_\mathrm{I}=0$ and the black line in (a), while the red circles mark the intersections between the yellow lines and the black line in (a).
\textbf{(e)} is to (b) as (d) is to (a), and \textbf{(f)} is to (e) as (c) is to (b) but along the $40~\unit{MHz}$ line.
\textbf{(g)} and \textbf{(h)} show the same data taken in the magnetic conditions of (e). The dashed lines in (h) correspond to the dashed lines in (e) of the same color. Dotted lines in (h) correspond to the dashed lines of the same color in (b) divided by an integer, representing multiphoton or higher harmonics signal. The red circles are calculated as in (e). Inset in (g) is a schematic representation of the spin flip-flop.
} 
\label{Fig:NVNV}
\end{center}
\end{figure*}
We observe dipolar spin flip-flop lines between ensembles of different \NVm orientations, specifically between the $T^{-1,0}_{\alpha}$ transition of the \NVm parallel to the main coil magnetic field and $T^{-1,+1}_{\beta,\gamma,\delta}$ transitions of the remaining three \NVm orientations [see Fig.~\ref{Fig:NVNV}(d)].  
The signal visibility of $T^{-1,+1}_{\beta,\gamma,\delta}$ is low as compared to $T^{-1,0}_{\alpha}$ [see Fig.~\ref{Fig:NVNV}(c)], and therefore the flip-flop ODMR signal is attributed mostly to the latter.  Applying magnetic field $B_\perp$ in the $\left[110\right]$ direction by increasing the current in the 2nd coil lifts the geometric degeneracy of the three non-parallel \NVm vectors [see Fig.~\ref{Fig:NVNV}(b)]. The visibility of $T^{-1,+1}_\beta$ is much smaller than the visibility of $T^{-1,+1}_{\gamma,\delta}$, probably due to the orientation with respect to the MW antenna, and the corresponding flip-flop signal was not detected. The relative strength of  the flip-flop signal [Fig.~\ref{Fig:NVNV}(f)] is comparable~\cite[p.~357]{abragam1982nuclear} to the same spin flip-flip case [Fig.~\ref{Fig:LAC} (b)] (i.e. both signals become detectable at an input power of $10~\unit{dBm}$). 

\begin{figure} \begin{center}
\includegraphics {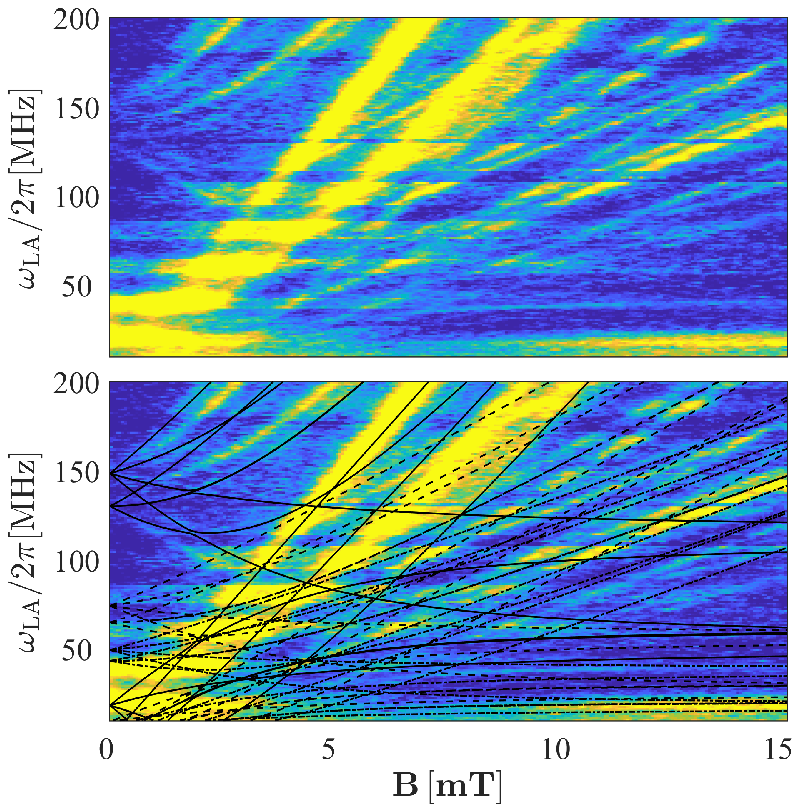}
\caption{P1 ODMR. Black solid/dashed/dot-dashed lines are some of the first/second/third order P1 transitions. 
The magnetic field orientation in the fit is the same as in Fig.~\ref{Fig:NVNV}, without current in the side coils.
}
\label{Fig:P1}
\end{center} \end{figure}

In addition, faint diagonal lines circa $53~\unit{mT}$ are attributed to a spin-flip process between the parallel \NVm and an additional  electronic spin-$1/2$ ensemble [Fig.~\ref{Fig:NVNV}(g)]. A suitable candidate is the P1 ensemble, though the signal strength does not allow reliable fit of the hyperfine lines. 
The expected P1 concentration is comparable to \NVm, yet the \NVm-P1 flip-flop is hardly registered.

A possible explanation is an efficient hyper-polarization of the P1 centers as compared to the non-parallel \NVm centers, leading to a lower flip-flop probability. The reduced non-parallel \NVm hyperpolarization may in turn be explained by spin flips during optical excitation-relaxation cycles or during photoionization and electron re-trapping. 
Yet a different explanation could be a photo-ionisation of the substitutional nitrogen, effectively reducing the concentration of the paramagnetic P1 centers~\cite{Doherty_041035,Lawson_6171}. However, the data of Fig.~\ref{Fig:P1} suggests that the P1 concentration is significant enough to alter the \NVm ODMR even when the P1 frequency is detuned far from the \NVm resonance.

\section{Additional Results}
\label{sec:add_res}
\begin{figure*}
\begin{center}
\includegraphics {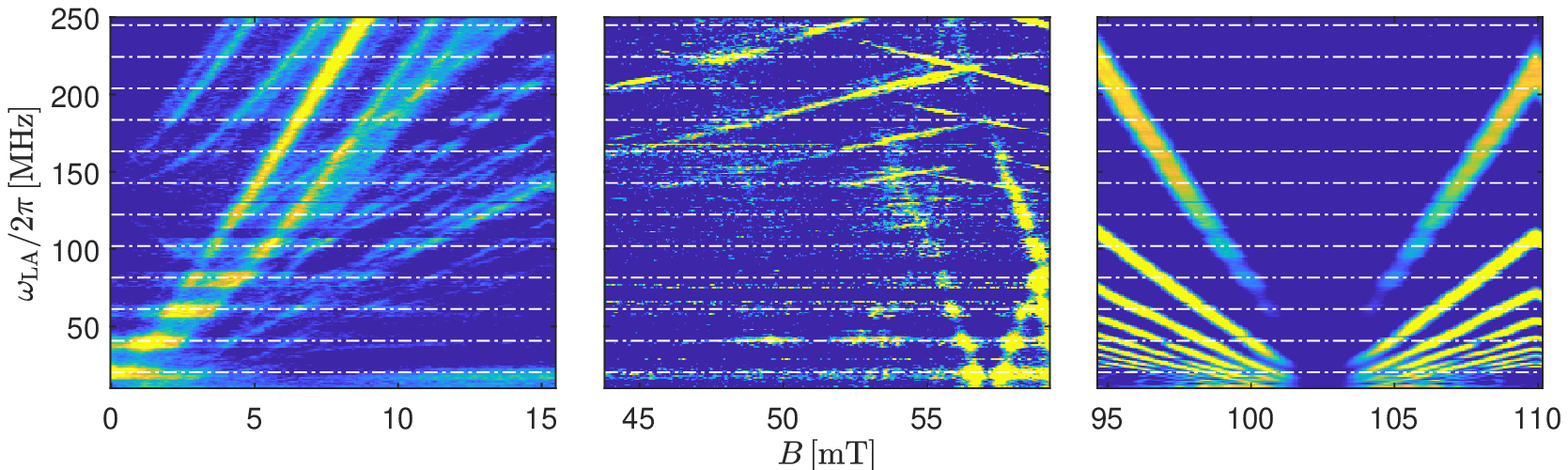}
\caption{Acoustic resonator. ODMR from Figs.~\ref{Fig:P1}, \ref{Fig:LAC} and \ref{Fig:NVNV}. White dashed lines plotted at $n\times f_\mathrm{a}=20.4~\unit{MHz}$ for integer $n$.}
\label{Fig:acoustic}
\end{center}
\end{figure*}

In low frequency region all observed ODMR lines are modulated with a pattern having a characteristic beating frequency given by $f_\mathrm{a}=20.4~\unit{MHz}$ (Fig.~\ref{Fig:acoustic}). The beating can be attributed to strain coupling to bulk acoustic standing waves in the diamond wafer \cite{Golter_143602,Nakamura_6405,Ovartchaiyapong2012}. However, the speed of sound of $2f_\mathrm{a}t=20.4~\unit{Km/s}$, which is derived from the measured beating frequency $f_\mathrm{a}$ and the thickness of the diamond wafer $t=0.5~\unit{mm}$, is about 13\%  higher than the values reported in ~\cite{Mcskimin_116,Mcskimin_2944,Nakamura_6405}.

\begin{figure} \begin{center}
\includegraphics {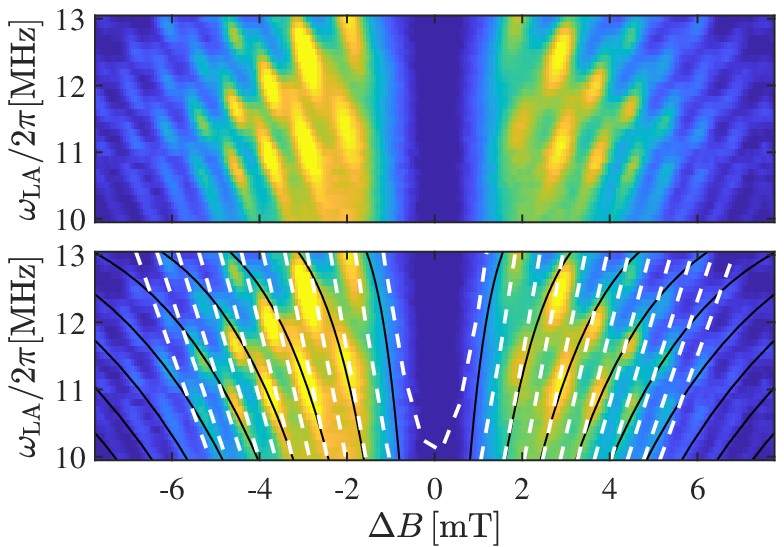}
\caption{Low frequency ODMR near GSLAC. The black lines represent empirical relation between frequency $f$ and magnetic field detuning from the GSLAC $\Delta B$ given by $f=f_\mathrm{arc}|\tanh(\Delta B/n B_\mathrm{arc})|$, with $f_\mathrm{arc}=13.9~\unit{MHz}$ ,$B_\mathrm{arc}=0.89~\unit{mT}$ and an integer $n$. White dashed lines are integer fractions of $\w[a]$, as calculated by Eq.~(\ref{eq:wa}) for the same $\theta$ as in Fig.~\ref{Fig:LAC}.} 
\label{Fig:arcs}
\end{center}
\end{figure}

The ODMR data shown in Fig.~\ref{Fig:arcs} for the frequency range of 10-13 MHz exhibits a complex modulation pattern. Three nuclear spins have been considered to explain the data. The nuclear spin 1 of nitrogen 14 of P1 defects is ruled out since its transition frequency is not expected to significantly change near \NVm GSLAC, whereas the data exhibits modulation pattern of arcs symmetric around $102~\unit{mT}$. The spin 1 of nitrogen 14 of the NV- defects is ruled out since its hyperfine coupling is too weak. First shell 13C has both sufficiently strong hyperfine coupling and its transition frequency strongly varies near $102~\unit{mT}$, however the high symmetry of the data is not compatible with 13C energy levels. Perplexedly, the arcs in the figure can be fitted by a single curve, stretched along the magnetic field axis by an integer factor (see caption of Fig.~\ref{Fig:arcs}).
\section{Discussion}\label{sec:discussion}
We report of optical detection of driven spin flip-flop, yet many questions remain unanswered with regard to the strength of perceived signal. For bath-control applications, of utter importance is the poor visibility of \NVm-P1 flip-flop line, which suggests high efficiency of P1 optical DNP process. With regards to \NVm-\NVm flip flops, the $\beta$ direction stands out, with both poor  visibility of $T^{-1,+1}_\beta$ transition and the lack of detectable flip-flop with $T^{-1,0}_\alpha$. In addition, no flip-flip lines at $2T^{-1,+1}_{\beta,\gamma,\delta}$ were detected. Further studies altering the relative orientations of the diamond, bias magnetic field and MW excitation may provide further insight regarding the strength of the various flip-flop lines, particularly in relation to the well understood flip-flip line strength. 
The strength of the flip-flop lines may be measured to optimize the efficiency of applied hyperpolarization protocols - i.e., after efficient hyperpolarization there are more aligned spins and less misaligned, hence flip-flop line is suppressed and the flip-flip line is emphasized. 

\section{Acknowledgments}
We thank Aharon Blank, Amit Finkler and Nir Bar Gill for fruitful discussions.  This work was supported by the Israel science
foundation, any by the Israel ministry of science.

\bibliographystyle{apsrev}
\bibliography{acompat,Eyal_Bib}

\end{document}